\documentclass[twocolumn,prd,nofootinbib,superscriptaddress,eqsecnum,tightenlines,8pt]{revtex4}

\usepackage{fancyhdr}
\usepackage{amsmath,amsfonts,amssymb}
\usepackage[colorlinks,citecolor=blue,linkcolor=blue,urlcolor=blue]{hyperref}

\usepackage{color}
\definecolor{red}{rgb}{1,0,0}
\definecolor{gre}{rgb}{0,0.6,0}
\definecolor{blu}{rgb}{0,0,1}

\usepackage[pdftex]{graphicx}
\usepackage{mathrsfs}
\DeclareGraphicsExtensions{.jpg, .JPG , .png , .pdf, .gif}
\def\be{\begin{equation}}
\def\ee{\end{equation}}

\renewcommand{\delta}{a}

\begin{document}

\title{Planck star phenomenology}

\date{\today}

\author{Aur\'elien Barrau}
\email{Aurelien.Barrau@cern.ch}
\affiliation{
Laboratoire de Physique Subatomique et de Cosmologie, Universit\'e Grenoble-Alpes, CNRS-IN2P3\\
53,avenue des Martyrs, 38026 Grenoble cedex, France\\
}%

\author{Carlo Rovelli}
\email{rovelli@cpt.univ-mrs.fr}
\affiliation{
Aix Marseille Universit\'e, CNRS, CPT, UMR 7332, 13288 Marseille, France\\
Universit\'e de Toulon, CNRS, CPT, UMR 7332, 83957 La Garde, France.\\
}%


\begin{abstract}
It is possible that black holes hide a core of Planckian density, sustained by quantum-gravitational pressure. 
As a black hole evaporates, the core remembers the initial mass and the final explosion occurs at macroscopic 
scale.  We investigate possible phenomenological consequences of this idea. Under several rough assumptions, we
estimate that up to several short gamma-ray bursts per day, around $10$ MeV, with isotropic distribution, can be expected 
coming from a  region of a few hundred light years around us.
\end{abstract}

\maketitle


\section{The model}

\begin{figure}
\centerline{\includegraphics[width=3.5cm]{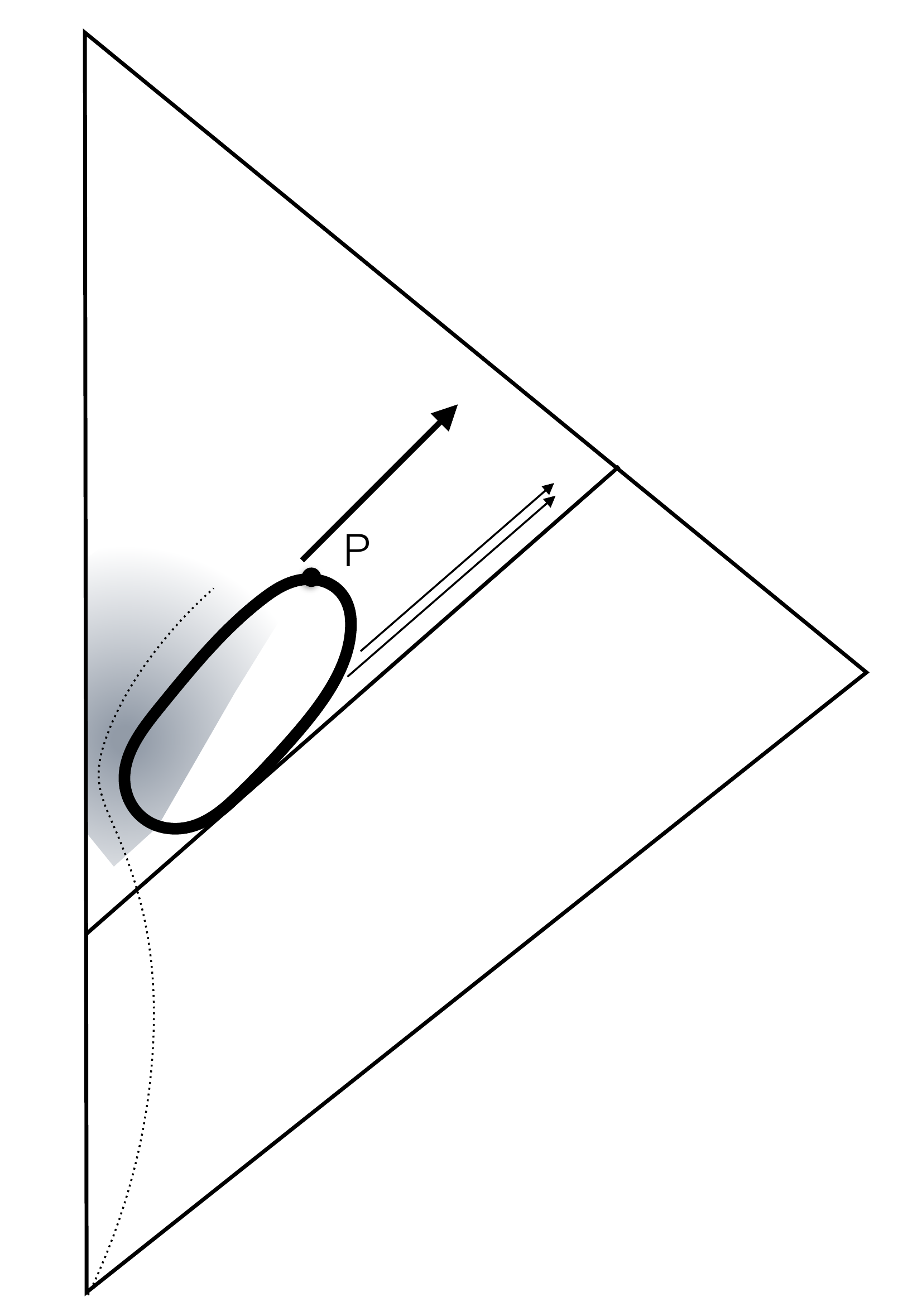}}
\caption{Penrose diagram of a collapsing star.  The dotted line is the external boundary of the star.  The shaded area is the region where quantum gravity plays an important role.  The dark line represents the two trapping horizons: the external evaporating one, and the internal expanding one. The lowest light-line is where the horizon of the black hole would be without evaporation.  $P$ is where the explosion happens.  The thin arrows indicate the Hawking radiation. The thick arrow is the signal studied in this paper.  }
\label{ps3}
\end{figure}

Recently, a new possible consequence of quantum gravity has been suggested \cite{Rovelli2014}. The idea is grounded in a robust result of loop cosmology \cite{Ashtekar2006}: when matter reaches Planck density, quantum gravity generates pressure sufficient to counterbalance weight. For a black hole, this implies that matter's collapse can be stopped before the central singularity is formed: the event horizon is replaced by a ``trapping'' horizon \cite{Ashtekar:2005cj} which resembles the standard horizon locally, but from which matter can eventually bounce out.  Because of the huge time dilation inside the gravitational potential well of the star, the bounce is seen in extreme slow motion from the outside, appearing as a nearly stationary black hole. The core, called ``Planck star", retains memory of the initial collapsed mass $m_i$ (because there is no reason for the metric of the core to be fully determined by the area of the external evaporating horizon) and the final exploding objects depends on $m_i$ and is much larger than Planckian \cite{Rovelli2014}. The process is illustrated by the conformal diagram of Fig. \ref{ps3}. 

In particular, primordial black holes exploding today may produce a distinctive signal. The observability of a quantum gravitational phenomenon is made possible by the amplification due to the large ratio of the black hole lifetime (Hubble time $t_H$) over the Planck time \cite{Amelino:13}. 

If this scenario is realised in nature, can the final explosion of a primordial Planck star be observed?  This is the question we investigate here. 

\section{Dynamics}

As a first step, we evaluate the energy of the particles emitted by the explosion of a primordial Planck star.  Let $m_f = \delta m_i$ be the final mass reached 
by the black hole before the dissipation of the horizon (at the point $P$ in Figure 1). In \cite{Rovelli2014}, an argument based on information conservation 
was given, pointing to the preferred value 
\be
\delta \sim \frac{1}{\sqrt{2}},
\label{m}
\ee
where $m_i$ is the initial mass. This value contradicts the expectation from the semiclassical
approximation; it follows from the hypothesis that the  black hole information paradoxes might be
resolved and firewalls avoided if the semiclassical approximation breaks down earlier than naively  expected,
as a consequence of the strong quantum gravitational effects in the core and the fact 
that they alter the effective causal structure of the evolving spacetime (the point $P$ 
is in the causal past of the quantum gravitational region.)  As shown by Hawking, 
non-rotating uncharged black holes emit particles with energy in the interval $(E,E+{\rm d}E)$ at a rate \cite{macgibbon}
\be
\frac{{\rm d^2N}}{{\rm d}E{\rm d}t}
=\frac{\Gamma_s}{h}\left[exp\left(\frac{8\pi G m E}{\hbar c^3}\right)-(-1)^{2s}\right]^{-1}
\label{hawk}
\ee
per state of angular momentum and spin $s$. The absorption coefficient $\Gamma_s$, that is the probability that the particle would be absorbed if it were incident in this state on the black hole, is a function of $E$, $m$ and $s$. By integrating this expression it is straightforward to show that the mass loss rate is given by
\be
\frac{{\rm dm}}{{\rm d}t}
=-\frac{f(m)}{m^2},
\label{loss}
\ee
where $f(m)$ accounts for the degrees of freedom of each emitted particle. As times goes on, the mass decreases, the temperature increases and new types of particles become ``available" to the black hole. Above each threshold, $f(m)$ is given approximately by  \cite{halzen}
\be
f(m)\approx(7.8\alpha_{s=1/2}+3.1\alpha_{s=1})\times 10^{24}~{\rm g}^3{\rm s}^{-1},
\label{fm}
\ee
where $\alpha_{s=1/2}$ and $\alpha_{s=1}$ are the number of degrees of freedom (including spin, charge and color) of the emitted particles. If $f(m)$ is assumed to be constant, {\it e.g.} $f(m)=f(m_i)$, the initial and final masses of a black hole reaching its final stage today, that is in a Hubble time $t_H$, are easy to calculate. In the Planck star hypothesis, the final stage is reached when $m=m_f\gg m_{Pl}$. Integrating Eq. (\ref{loss}) leads to:
\be
m_i=\left(\frac{3t_Hf(m_i)}{1-\delta^3}\right)^\frac{1}{3}.
\label{mi}
\ee

In practice, to account for the smooth evolution of $f(m)$ when the different degrees of freedom open up, a numerical integration has to be carried out. This leads, for $\delta=1/\sqrt{2}$, to 
\be
m_i\approx6.1\times 10^{14}~{\rm g},
\label{minum}
\ee
and
\be
m_f\approx4.3\times 10^{14}~{\rm g}.
\label{mfnum}
\ee
The value of $m_i$ is very close to the usual value $m_*$  corresponding to black holes requiring just the age of the Universe to fully evaporate. This was expected as the process is explosive. The details may  change depending on the exact shape chosen for $f(m)$. 

The value of the radius when $m$ reaches $m_f$ is $r_f\approx 6.4\times 10^{-14}~{\rm cm}$. The size of the black hole is the only scale in the problem and therefore fixes the energy of the emitted particles in this last stage. We assume that all fundamental particles are emitted with the same energy taken at 
\be
E_{burst}=hc/(2r_f)\approx 3.9 ~{\rm GeV}.
\ee 
Of course, a more reliable model would be desirable but this is the most natural hypothesis at this stage.\\

\section{Single event detection}

\begin{figure}
\centerline{\includegraphics[width=9cm]{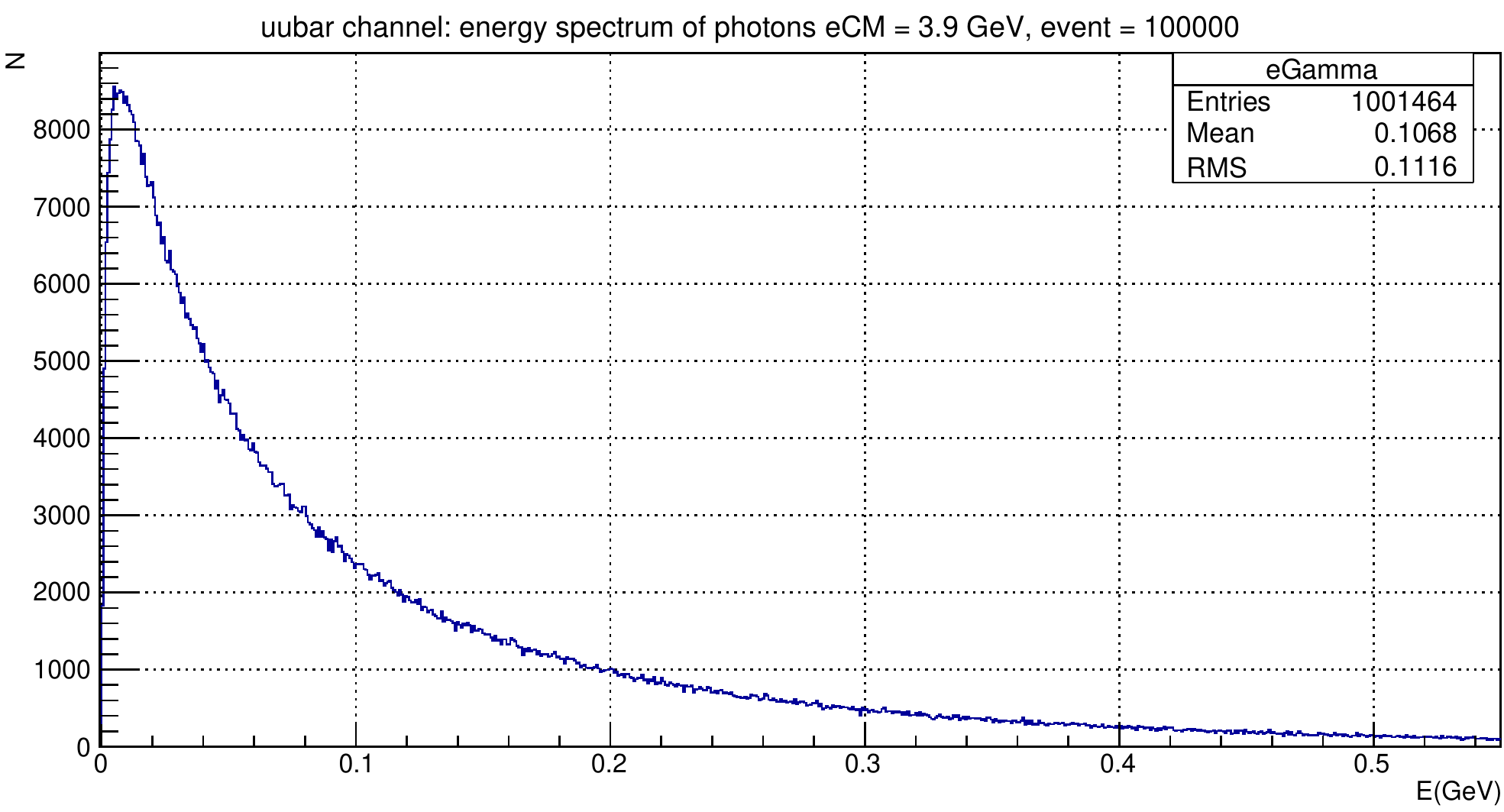}}
\caption{Gamma-ray spectrum resulting from $10^5$ $u\bar{u}$ jets (linear scales).}
\label{quark}
\end{figure}

We now study the signal that evaporating Planck star would produce. From the phenomenological viewpoint, it is natural to focus on emitted gamma-rays: charged particles undergo a diffusion process in the stochastic magnetic fields and cannot be used to identify a single event whereas neutrinos are hard to detect. The important fact is that most of the emitted gammas are not emitted at the energy $E_{burst}$. Only those directly emitted will have this energy. But assuming that the branching ratios are controlled, as in the Hawking process, by the internal degrees of freedom, this represents only a small fraction (1/34 of the emitted particles). Most gamma-rays will come from the decay of hadrons produced in the jets of quarks, notably from neutral pions. $E_{burst}$ is already much smaller than the Planck scale but the mean emission of emitted photons is even smaller.

To simulate this process, we have used the "Lund Monte Carlo" PYTHIA code (with some scaling approximations due the unusually low energy required for this analysis). It contains theory and models for a number of physics aspects, including hard and soft interactions, parton distributions, initial- and final-state parton showers, multiple interactions, fragmentation and decay. PYTHIA allowed us to generate the mean spectrum expected for secondary gamma-rays emitted by a Planck star reaching the end of its life. The main point to notice is that the mean energy is of the order of $0.03\times E_{burst}$, that is in the tens of MeV range rather than in the GeV range. In addition, the multiplicity is quite high at around 10 photons per $q\bar{q}$ jet. Fig. \ref{quark} shows the mean spectrum of photons resulting from $10^5$ jets of 3.9 GeV $u\bar{u}$ quarks.\\

\begin{figure}
\centerline{\includegraphics[width=9.3cm]{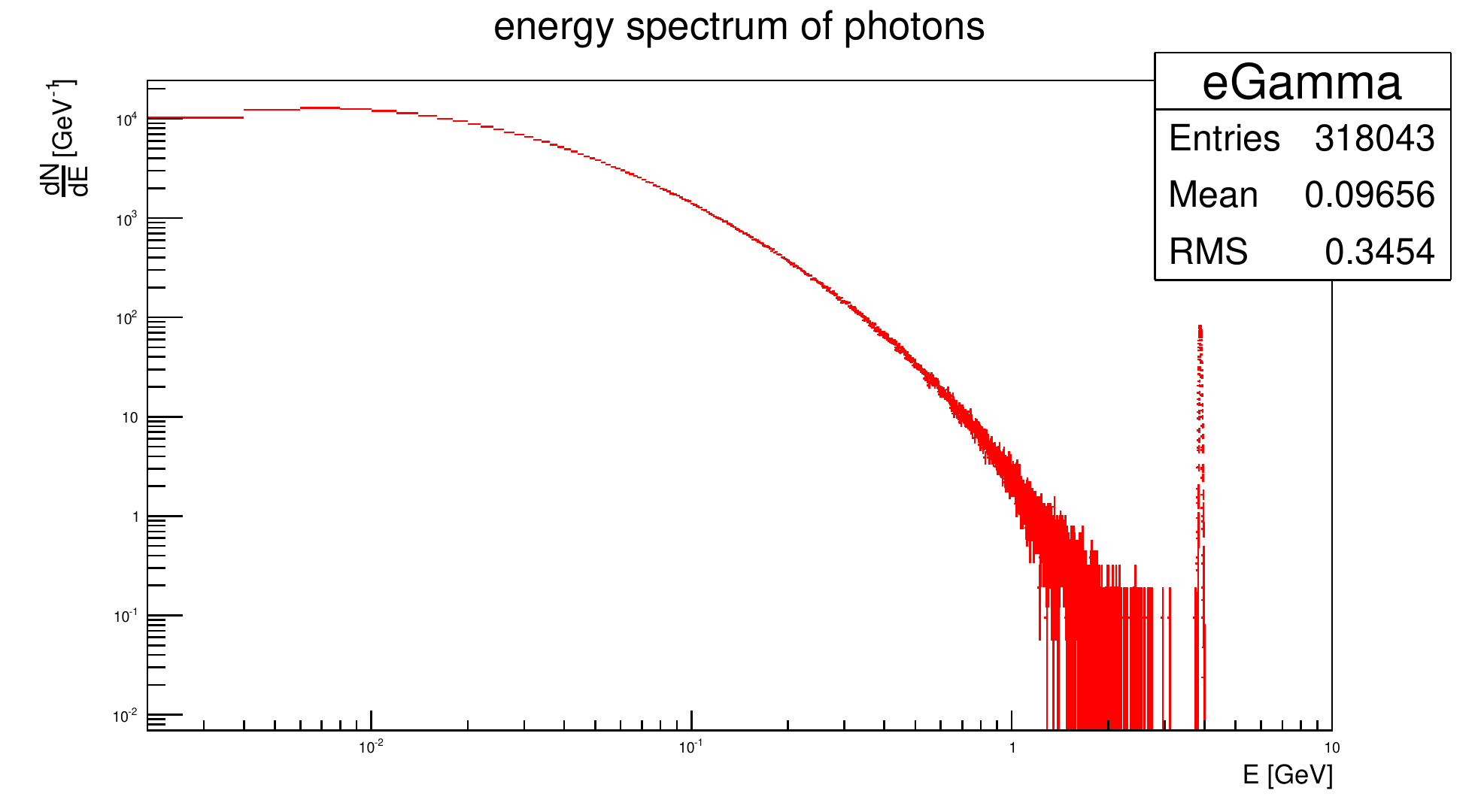}}
\caption{Full spectrum of gamma-rays emitted by a decaying Planck star (log scales).}
\label{full}
\end{figure}

It is straightforward to estimate the total number of particles emitted $m_f/E_{burst}$ and then the number of photons $<N_{burst}>$  emitted during the burst. As for a black hole radiating  by the Hawking mechanism, we assume that the particles emitted during the bursts (that is those with $m<E_{burst}$) are emitted proportionally to their number of internal degrees of freedom : gravity is democratic. The spectrum resulting from the emitted $u,d,c,s$ quarks ($t$ and $b$ are too heavy), gluons and photons is shown on Fig. \ref{full}. The little peak on the right corresponds to directly emitted photons that are clearly sub-dominant. By also taking into account the emission of neutrinos and leptons of all three families (leading to virtually no gamma-rays and therefore being here a pure missing energy), we obtain $<N_{burst}>\approx4.7\times 10^{38}$. \\

The question of the maximum distance at which a single burst can be detected naturally arises. If one requires to measure $N_{mes}$ photons in a detector of surface $S$, this is simply given by
\be
R_{det}=\sqrt{\frac{S<N_{burst}>}{4\pi N_{mes}}}.
\label{dist}
\ee

If we set, {\it e.g.}, $N_{mes}\approx 10$ photons in a 1 m$^2$ detector, this leads to $R\approx 205$ light-years. Otherwise stated, the ``single event" detection of exploding Planck stars is {\it local}. The maximum distance at which such an event can be detected is just a few tens times the distance to the nearest star. This is a tiny galactic patch around us.\\

This already has an interesting consequence. Except if there is a significant dark matter clump within this small sphere -- which is unlikely--, the signal is expected to be {\it isotropic} as the halo is homogeneous at this scale. This is very different from most galactic signals usually peaked either in the galactic center direction, or the direction of the motion of the solar system for directed searches for dark matter. This means that the Planck star signal could mimic a cosmological origin.\\

Let us call the local density of dark matter $\rho^{DM}_* \approx 0.3 \pm 0.1$ GeV/cm$^3$ \cite{bovy}. If Planck stars reaching $m_f$ were to saturate the dark matter bound, and if they cluster as ordinary cold dark matter, their number within the detectable horizon would be 
\be
N_{det}^{max}=\frac{4\pi \rho^{DM}_*}{3m_f}\left(\frac{S<N_{burst}>}{4\pi N_{mes}}\right)^\frac{3}{2}\approx 3.8 \times 10^{22}.
\label{dist}
\ee
However, as their history is not that different from the one of standard primordial black holes with the same initial mass (of course the end of their lives {\it is} different but the total energy emitted remains the same) the usual constraint $\Omega^{PBH}<10^{-8}$ for initial masses around $10^{15}$ g basically holds \cite{jane}. This leads to 
\be
N_{det}<\frac{4\pi \rho^{DM}_*\Omega^{PBH}}{3m_f}\left(\frac{S<N_{burst}>}{4\pi N_{mes}}\right)^\frac{3}{2} \approx 3.8 \times 10^{14} .
\label{dist}
\ee
This number is still quite high. It shows that the individual detection is far from being, in principle, out of reach. \\

One can then estimate the number of events that can be expected for a given observation time $\Delta t$. This corresponds to Planck stars that have masses between $m_f$ and $m(\Delta t)$ at the beginning of the observation time, within the volume $R<R_{det}$. In this case, $m(\Delta t)$ is simply:
\be
m(\Delta t)=\left(m_f^3+3f(m)\Delta t\right)^\frac{1}{3}. 
\label{mdeltat}
\ee

This number $n(\Delta t)$ is (estimated for a unit volume) given by
\be
n(\Delta t)=\int_{m_f}^{m(\Delta t)}\frac{{\rm d}n}{{\rm d}m}{\rm d}m,
\label{dist}
\ee
where ${\rm d}n/{\rm d}m$ is the differential mass spectrum of Planck stars today, still per unit volume. Importantly, the shape of this mass spectrum in the interesting region is mostly independent of the initial shape. This is exactly true only in the limit $m\ll 3f(m)t_H$ and constitutes a rough approximation here. To get orders of magnitude, we however assume this to be correct. In this case, due to the dynamics of the evaporation, ${\rm d}n/{\rm d}m \propto m^2$. This can be straightforwardly seen by writing ${\rm d}n/{\rm d}m={\rm d}n/{\rm d}m_i\times {\rm d}m_i/{\rm d}m$, where ${\rm d}n/{\rm d}m_i$ is the initial mass spectrum and ${\rm d}m_i/{\rm d}m=m^2(3f(m)t+m^3)^{-2/3}$.\\

If primordial black holes leading to Planck stars are formed through a kind of phase transition in the early universe, their mass spectrum can be very narrow. In the most extreme optimistic case, this would cover exactly the range of masses reaching $m_f$ in the observation time window $\Delta t$. In that case, the number of observed explosions would be $N_{expl}=N_{det}$. This is obviously unrealistic.  On the other extreme, one can assume a very wide mass spectrum. As the primordial cosmological power spectrum $P(k)\propto k^n$ is now known to be red $(n<1)$ whereas it would have had to be blue to produce a sizable amount of primordial black holes by standard processes, the usual historical spectrum
\be
\frac{{\rm d}n}{{\rm d}m_i}=\alpha m_i^{-1-\frac{1+3w}{1+w}},
\label{spec}
\ee
where $w=p/\rho$ is the equation of state of the Universe at the formation epoch, can only be taken as an approximation on a reduced mass interval. This is however not unrealistic in models like Starobinski's broken scale invariance.

This leads, assuming that the formation occurred in the radiation dominated era, to a contemporary spectrum reading
\be
\frac{{\rm d}n}{{\rm d}m}\sim\alpha\left[m^{-\frac{5}{2}}\Theta(m-m_*)+m_*^{-\frac{9}{2}}m^2\Theta(m_*-m)\right].
\label{spectoday}
\ee

The number of expected "events" during an observing time $\Delta t$ is given by
\be
N(\Delta t)=\frac{\int_{m_f}^{m(\Delta t)}\frac{{\rm d}n}{{\rm d}m}{\rm d}m}{\int_{m_f}^{m_{max}}\frac{{\rm d}n}{{\rm d}m}{\rm d}m}\Omega^{PBH}N_{det}^{max}\Omega_{sr},
\label{evts}
\ee
where $m_{max}$ is the maximum mass up to which we assume the mass spectrum to be given by Eq. (\ref{spectoday}) and $\Omega_{sr}$ is the solid angle acceptance of the considered detector. Here, the value of $\Omega^{PBH}$ is not easy to constrain. An upper limit can be taken conservatively at $10^{-8}$. To fix orders of magnitude, if we set $m_{max}=m_*$ and a density of a few percents of the maximum allowed density, that is $\Omega^{PBH}\sim 10^{-10}$, this leads to one event per day. Detection is not hopeless.

\section{Very short gamma-ray bursts and diffuse emission}

Could it be that those events have already been detected? In particular : could they be associated with some gamma-ray bursts (GRBs)? There are two main classes of GRBs : the long ones and the short ones. The long ones are quite well understood and are associated with the deaths of massive stars and have no link with this study. Our model for the end of Planck stars does not allow for the calculation of any kind of light curve. This would be far beyond the current development. But, obviously, the time-scale is expected to be short.\\

If Planck star explosions are to be associated with some of the measured GRBs, this would obviously be with short gamma-ray bursts (SGRBs) \cite{nakar}. This could make sense for several reasons. Firstly, SGRBs are the less well understood. In particular, the redshifts are not measured for a large fraction of them, which means that they could, in principle, be of local origin. Secondly, SGRBs are known to have a harder spectrum and some of them do indeed reach the energies that we have estimated in this work. Thirdly, a sub-class of SGRB, the very short gamma ray bursts (VSGRBs), do exhibit an even harder spectrum and can be assumed to originate from a different mechanism as the SGRB time distribution seems to be bimodal \cite{cline}.\\

Nothing can be concluded at this stage but it is worth noticing that there are indications that some VSGRBs are compatible with a Planck star origin. The fact that their angular distribution favors a cosmological origin is in fact also fully compatible with a local bubble origin. \\

\begin{figure}
\centerline{\includegraphics[width=9.3cm]{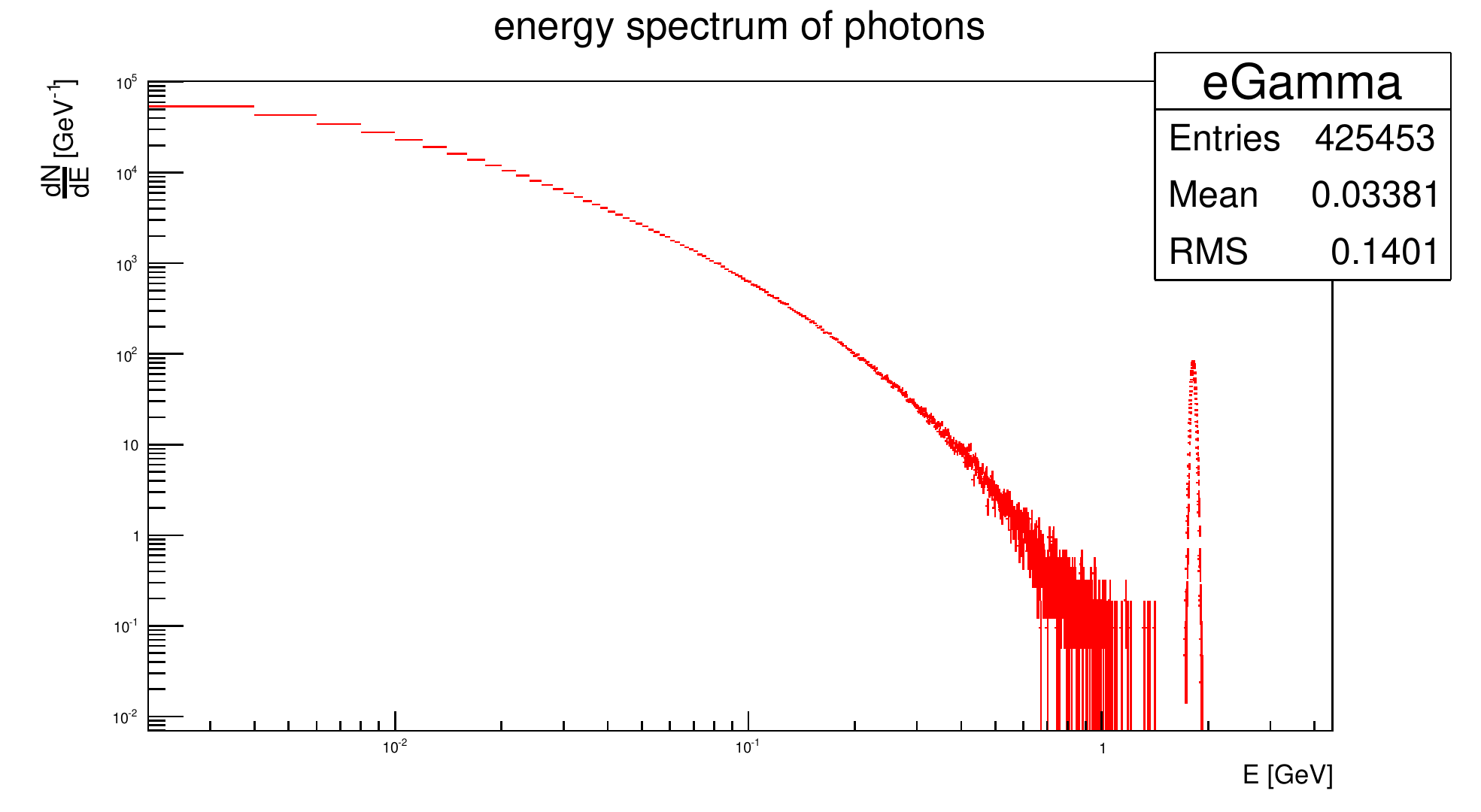}}
\caption{Full spectrum of gamma-rays emitted by a decaying Planck star at $z=3$ (log scales).}
\label{plotz3}
\end{figure}

Another important question to address is the possible diffuse emission from Planck stars. Not the single event detection but the integrated signal over huge distances. In that case, no time signature can be expected but the global spectrum can retain some specific characteristics. Two effects are competing. Let us consider a Planck star exploding at a redshift $z$. Because, to be detected now, it has exploded in the past, it means that the amount of cosmic time required for it to reach its final mass was smaller than in our vicinity. Its initial and final masses were therefore smaller. This means that the mean energy of emitted particles was {\it higher} (and the total amount smaller) than for Planck stars exploding within the Galaxy. But just because of the redshift, the measured energy is {\it smaller} than the emitted one by a factor $(1+z)$. In practice, this second effect slightly dominates over the first one. For example, a Planck star exploding at $z=3$ emits photons with a mean energy higher than around us by a factor 1.9. This energy is then redshifted by a factor 4. Fig. \ref{plotz3} shows the same spectrum than in Fig. \ref{full} but for a $z=3$ Planck star (it is not just a rescaling of the previous one as the energy of jets changes). \\

When considering a diffuse signal with very small absorption effects, like gamma-rays in this range, the integration effect is drastic. For each shell, the number of detected photons is inversely proportional to the squared shell distance because of the solid angle effect. But the number of sources per shell is proportional to the squared distance. The two effects compensate (modulo the slight energy variation mentioned in the previous paragraph) and each shell contributes the same. The neat flux is therefore fully determined by the cutoff. We leave the detailed study for future works as, in this case, it would be mandatory to take into account not only the explosion spectrum but also the last stages of the Hawking spectrum due to the evaporation before the explosion. It can however be easily concluded that the signal can in principal be detected as the mean density required to produced a sizable amount of gamma-rays is very small. 

\section{Conclusion and prospects}

We have shown that the detection of individual explosions of Planck stars is not impossible and we have established the main spectral characteristic of the signal. Quantum gravity might  show up in the tens of MeV range. We have  estimated the order of magnitudes for the expected frequency of events: in some cases, it might not be small.  Several approximations about the dynamics of the evaporation can be improved. The details of the explosion remain to be investigated. The shape of the diffuse integrated signal, and more specifically its potential specific signature allowing to distinguish it from standard primordial black holes, requires a full numerical analysis. It could also be interesting to investigate the emission of charged articles, in particular positrons and antiprotons (some interesting threshold effets could be expected). As well known, the much smaller horizon can be compensated by the large galactic confinement effect.

\acknowledgments
A.B. would like to thank Wei Xiang.

\end{document}